\documentclass{Interspeech}

\interspeechcameraready

\usepackage{comment}

\usepackage{tabularx}
\usepackage{graphicx}
\usepackage{adjustbox}
\usepackage{multicol,multirow}
\usepackage{hyperref}
\title{Are audio DeepFake detection models polyglots?}

\author[affiliation={1}]{Bartłomiej}{Marek}
\author[affiliation={2}]{Piotr}{Kawa}
\author[affiliation={2}]{Piotr}{Syga}

\affiliation{}{CISPA – Helmholtz Center for Information Security}{Germany}
\affiliation{}{Wrocław University of Science and Technology}{Poland}

\address{
  $^1$Wrocław University of Science and Technology, Poland \\
  $^2$CISPA – Helmholtz Center for Information Security, Germany}
\email{bartlomiej.marek@cispa.de, \{piotr.kawa, piotr.syga\}@pwr.edu.pl}

\begin{document}

\maketitle
\keywords{Audio DeepFakes, DeepFake detection, multilingual audio DeepFakes}

\begin{abstract}
Since the majority of audio DeepFake (DF) detection methods are trained on English-centric datasets, their applicability to non-English languages remains largely unexplored. 
In this work, we introduce a benchmark for the multilingual audio DF detection challenge by evaluating various adaptation strategies. Our experiments focus on analyzing models trained on English benchmark datasets, as well as intra-linguistic (same-language) and cross-linguistic adaptation approaches. Our results indicate considerable variations in detection efficacy, highlighting the difficulties of multilingual settings. We show that limiting the training dataset to English negatively impacts the efficacy, while using even a small amount of data in the target language proves more beneficial for detection than adding larger volumes of data from multiple non-target languages combined.
\end{abstract}

\section{Introduction}
The rapid growth of generative AI, especially in voice synthesis, has made it easier to create personalized voices. Technologies such as text-to-speech (TTS) and voice cloning (VC) need only seconds of voice input to produce convincing replicas~\cite{zhang2023surveyaudiodiffusionmodels}. While useful for personal assistants, these tools can also be misused, such as for audio DeepFakes (DF). Malicious DFs can undermine media credibility and enable harmful manipulation, from political disinformation such as AI-generated Polish campaign ads~\footnote{\href{https://notesfrompoland.com/2023/08/25/opposition-criticised-for-using-ai-generated-deepfake-voice-of-pm-in-polish-election-ad/}{https://notesfrompoland.com/2023/08/25/opposition-criticised-for-using-ai-generated-deepfake-voice-of-pm-in-polish-election-ad}} to financial fraud, such as a \$25.6M scam in Hong Kong using executive impersonation~\footnote{\href{https://edition.cnn.com/2024/02/04/asia/deepfake-cfo-scam-hong-kong-intl-hnk/index.html}{https://edition.cnn.com/2024/02/04/asia/deepfake-cfo-scam-hong-kong-intl-hnk/index.html}}.

Despite extensive research on DeepFake detection, a challenge similar to spoofing countermeasures~\cite{DBLP:journals/corr/abs-2111-07725}, key issues persist, including limited diverse data and poor generalization~\cite{muller2022does}. Recent advancements~\cite{casanova2024xttsmassivelymultilingualzeroshot,sharma2020fastgriffinlimbased,kim2021conditionalvariationalautoencoderadversarial} have democratized voice technology, enabling users to generate high-quality speech across languages. A Recorded Future Inc. report~\footnote{\href{https://go.recordedfuture.com/hubfs/reports/ta-2024-0924.pdf}{https://go.recordedfuture.com/hubfs/reports/ta-2024-0924.pdf}} found 82 DeepFakes of public figures in 38 countries (July 2023–July 2024), with 30 of them holding elections, underscoring DFs' global impact. Mitigation may require localized strategies, yet most models remain trained primarily on English and Chinese due to dataset availability.

The recent Multi-Language Audio Anti-Spoofing Dataset (MLAAD)~\cite{muller2024mlaadmultilanguageaudioantispoofing} enables research on cross-language model generalization. 
Our work introduces a benchmark to evaluate multilingual DF detection, focusing on adaptation strategies: fine-tuning, training from scratch, or using English pre-trained models to enhance performance across languages. 

Our work extends prior studies~\cite{chetia-phukan-etal-2024-heterogeneity, liu2024towards} by adding intra- and cross-lingual adaptations, language correlations, and investigating language families while limiting potential biases from overlapped synthesizers~\cite{trident}.
Specifically, we take a broader perspective by analyzing languages from three families: Germanic (English, German), Romance (French, Italian, Spanish), and Slavic (Polish, Russian, Ukrainian), with a focus on designing a benchmark that is as unbiased as possible using public data.

We aim to determine whether language-specific data are necessary for accurate detection and how to best adapt English-trained models when only a small dataset in the target language is available. We empirically explore three essential research questions (RQ) in this area. Specifically, our objective is to check to what extent the detection efficacy varies by language, whether English benchmark-trained models are sufficient for effective cross-linguistic detection, and which targeted strategies best support DF detection in specific languages, precisely intra- or cross-lingual adaptations, \textbf{even assuming access to very limited resources in a specific language}.

\noindent\textbf{RQ1: \textit{
To what extent can English-trained detection DFs models generalize to multilingual scenarios? }}\\
\noindent Current, publicly available benchmarks are English-centric, potentially leaving detection models underprepared for real-world scenarios involving non-English audio. This explores the challenges of relying solely on benchmark-trained models for diverse linguistic contexts, focusing on their performance in specific non-English languages.

\noindent\textbf{RQ2: \textit{ How does language choice influence DeepFake detection effectiveness?}} \\
\noindent Despite advances in audio DF detection, there is limited understanding of how language influences detection efficacy. This analyzes how detection effectiveness varies across languages and whether multilingual training data enhance or degrade performance compared to language-specific approaches.

\noindent\textbf{RQ3: \textit{Which is the more effective adaptation strategy: language-specific with limited data, or multilingual with larger datasets?}} \\
\noindent  We compare training a language-specific model with limited data, using an English-trained model, and fine-tuning it with language-specific or multilingual data.  We aim to determine if a more targeted or diverse language dataset would be more effective for our strategy. In the latter case, it would be necessary to determine which languages are most suitable for adaptation.

\noindent The codebase related to our research can be found in \footnote{\href{https://github.com/bartlomiejmarek/are\_audio\_df\_polyglots}{https://github.com/bartlomiejmarek/are\_audio\_df\_polyglots}}.

\begin{table*}[ht!]
\caption{Hours of training data used for W2V+AASIST XLS-R 300m~\cite{babu2021xls} and Whisper medium~\cite{whisper}.}

    \label{tab:train-data}
    \centering
    \resizebox{0.6\textwidth}{!}{
    \begin{tabular}{c|cccccccc}
    \hline
    \multirow{2}{*}{\textbf{Model}} &  \multicolumn{8}{c}{\textbf{Languages}} \\
    \cline{2-9}
     &  de & en & es & fr & it & pl & ru & uk \\
    \hline
    \midrule
    W2V XLS-R 300m    & 69\,493 & 25\,378 & 22\,258 & 23\,973 & 21\,943 & 20\,912 & 166 & 72 \\
    Whisper & 438\,218 & 13\,344 & 11\,100 & 9\,752 & 2\,585 & 4\,278 & 9\,761 & 697 \\
    \hline
    \end{tabular}
    }
\end{table*}

\section{Related works}
\label{sec:relatedworks}
DeepFake detection has gained attention as TTS and VC algorithms create increasingly realistic audio fakes. A substantial portion of the research in this domain has used ASVspoof 
datasets~\cite{wu15e_interspeech,asvspoof2019,yamagishi21_asvspoof,ASVSpoof5}, considered a gold standard for the anti-spoofing domain. However, these datasets do not fully represent real-world scenarios in terms of language coverage, as they exclusively consist of English samples.

The discrepancy between generation methods in the training and test sets leads to significant performance drops, especially in real-world settings. While~\cite{kawa2022attack} proposed a method for dealing with unseen methods of DF generation,~\cite{muller2022does} showed that detection models have even greater difficulty in the correct classification of real-world samples, publishing the "In the Wild" dataset.
Moreover, most published datasets contain only English samples~\cite{Wavefake,khalid2021fakeavceleb,av-df-1m}, with~\cite{Wavefake} including some Japanese utterances and~\cite{ADD,FMFCC,HAD,CFAD} providing Chinese ones.

Recent publications indicate that the effectiveness of detection methods significantly degrades across linguistic boundaries~\cite{10744454, huangspeechfake}. 
The results suggest that detection systems exhibit substantial bias toward training language characteristics. Specifically, performance drops significantly when evaluated using unfamiliar languages or accents (even within the same language).~\cite{10744454}.
On the other hand, the performance of detection methods is also heavily dependent on the overlap between the speech generators seen during training and those used in testing, thus highlighting the inability of existing models to generalize to unseen synthesis techniques~\cite{trident, huangspeechfake}.

A recent Multi-Language Audio Anti-spoofing Dataset (MLAAD)~\cite{muller2024mlaadmultilanguageaudioantispoofing} attempts to address the gap, providing over 76,000 utterances in 23 languages, with DF generated using 54 systems to present samples with varied distributions. This publicly available, large-scale audio fake corpus spans over 160 hours and forms a complete dataset with the {M-AILABS} Speech Dataset, consisting of authentic audio recordings from public domain books.
For this research, we used the third version of the MLAAD dataset
\footnote{\href{https://owncloud.fraunhofer.de/index.php/s/tL2Y1FKrWiX4ZtP}{https://owncloud.fraunhofer.de/index.php/s/tL2Y1FKrWiX4ZtP}}.

Notably, due to the large number of languages in the dataset, the number of samples in each language is limited.
Given the problem with the generalization of the models, we cannot be sure if the efficacies of the detection models rely on the target language, i.e., the language in which the utterance is spoken. This motivated this paper so that further research could focus on the most promising way to detect non-English DFs. 

\section{Experimental setup}
\label{sec:experimental}
Throughout the article, we investigate the efficacy of English-trained detection models in detecting DFs for various languages and provide a strategy to improve the detection efficacy for DF utterances in languages with severely limited datasets that might be used to train or fine-tune the detection model. In this study, we concentrate on the languages represented in the M-AILABS Speech Dataset, which provides a comprehensive sample of authentic language data in these languages. We have selected English (en) and German (de) from the Germanic family, French (fr), Italian (it), and Spanish (es), which represent the Romance languages, and Polish (pl), Russian (ru), and Ukrainian (uk), which represent the Slavic languages. 

\subsection{Models}

Audio DF detection methods are based on either direct waveform analysis via end-to-end models~\cite{rawgat-st, jung2020improved},  or 
feature-based methods employing front-end extractors, which derive acoustic properties from raw audio for subsequent deep-learning~\cite{DBLP:conf/odyssey/TakTWJYE22, DBLP:conf/wifs/AfcharNYE18, kawa2023improved}. 
Our research systematically examines the leading DF detection models~\cite{rawgat-st, DBLP:conf/wifs/AfcharNYE18, DBLP:conf/odyssey/TakTWJYE22, DBLP:journals/corr/abs-2110-01200}, assuming limited availability of non-English resources. Specifically, we evaluate the following scenarios:
i) English-trained detection model on the target language,
ii) training from scratch directly on the target language,
iii) fine-tuning pre-trained English models on target languages,
iv) fine-tuning with a single related or unrelated language, 
v) fine-tuning with multiple languages while excluding the target language. 
For all the experiments, we assume that non-English resources are severely limited. 
Given that two of our models utilize SSL architectures, we selected a pretrained W2V XLS-R 300m~\cite{babu2021xls} and Whisper medium~\cite{whisper}. In Table~\ref{tab:train-data} we can observe the amount of data used to train the specific language.

\subsection{Datasets} We establish a baseline by utilizing models trained on the entire English benchmark dataset ASVspoof2019 LA, consisting of the training, development, and evaluation sets, and treat them as a reference point for language-specific fine-tuning, as most researched audio DFs detection solutions used English or Chinese for training (some of the latter's sets, e.g., ADD~\cite{ADD}, were not publicly available during this research). 

\begin{table*}[ht!]
\centering
\caption{Edit‑distance (ED) means $\pm$ standard deviations for fake and original transcriptions for MLAAD dataset~\cite{muller2024mlaadmultilanguageaudioantispoofing}.}\resizebox{0.7\textwidth}{!}{%
\begin{tabular}{l|cccccccc}
    \hline
                 & de & en & es & fr & it & pl & ru & uk \\ \midrule
                     \hline
$\text{ED}_{\text{fake}}$    & 2.4 $\pm$2.6 & 4.5 $\pm$4.9   & 2.5 $\pm$3.1  & 4.8 $\pm$5.1  & 9.5 $\pm$10.4 & 5.8 $\pm$14.0 & 4.9 $\pm$4.6  & 23.2 $\pm$34.0 \\
$\text{ED}_{\text{original}}$ & 3.5 $\pm$7.4 & 2.8 $\pm$3.7    & 2.1 $\pm$2.5  & 4.0 $\pm$4.3  & 2.0 $\pm$2.0  & 2.7 $\pm$3.1  & 2.9 $\pm$3.6  & 6.1 $\pm$11.3 \\ \bottomrule
\end{tabular}}

\label{tab:edit-distance}
\end{table*}
Several languages of MLAADv3 are generated using four architectures (\textit{XTTS v1.1}, \textit{XTTS v2}, \textit{Griffin-Lim}, \textit{VITS})~\cite{muller2024mlaadmultilanguageaudioantispoofing}.  To avoid any overlap between fine-tuning and evaluation architectures, we employ strict constraints.
We perform fine-tuning of the models, initially pre-trained on ASVspoof 2019, with audio samples generated using \textit{VITS} and \textit{Griffin-Lim}, and later evaluate on spoof samples synthesized with \textit{XTTS v1.1} and \textit{XTTS v2}. To ensure fairness, we also reverse this setup, 
(\textit{XTTS} architectures are used for fine-tuning, whereas \textit{VITS} and \textit{Griffin-Lim} for testing). This guarantees no overlap between the methods in fine-tuning and test datasets. As shown recently~\cite{trident}, such overlap could lead to incorrect conclusions on language transferability because the ease or challenge might stem from generator-specific artifacts rather than linguistic properties. This separation across samples is essential to maintain the integrity of our comparisons. 
For evaluation assessment, we calculate the Equal Error Rate (EER) (in \%) for both runs separately and report an average of them. 
Our approach focuses on conducting standardized comparisons within the available data resources.

Even for this limited number of languages and architectures, we took proactive steps to generate missing data to avoid potential bias. Specifically, we synthesized samples using \textit{XTTS v1.1} and \textit{XTTS v2} for English and \textit{VITS} samples for Russian, which were not included in the original MLAAD. 
Moreover, it is important to note that the original MLAAD dataset does not contain Ukrainian samples generated using \textit{XTTS v1.1} and \textit{XTTS v2} due to the limitations of these generators. This led us to use \textit{GlowTTS} and \textit{Facebook Massively Multilingual Speech} instead of excluding this language from our study.

\subsection{Hyperparameters}
We train and fine-tune the models utilizing distinct hyperparameter configurations depending on the model architecture. For the lightweight architectures (LFCC+AASIST, LFCC+MesoNet, RawGAT-ST), we utilize a learning rate of 5e-03 and a weight decay of 2.5e-05. In contrast, for the larger models, W2V+AASIST and Whisper+AASIST, we utilize a learning rate of 5.0e-06 and a weight decay of 5e-07. Due to limited data resources, we use an unchanged hyperparameter configuration, as well as a more traditional, lower learning rate and weight decay typically applied during fine-tuning.
Thus, for fine-tuning lightweight models, we use a learning rate of 1e-04,
a weight decay of 5.0e-06, an learning rate of 5.0e-06, and a weight decay of 5.0e-07.  Similarly, we fine-tune W2V+AASIST and Whisper+AASIST using either a learning rate of 2.5e-6 or 5e-6, both with a weight decay of 2.5e-7. Furthermore, the SSL front-ends (W2V and Whisper) remain fully trainable during both training and fine-tuning, with all weights unfrozen. We do not apply any data augmentation techniques.
The evaluation results are the mean of 10 runs of 90\% of the test dataset.

\section{Results}
\label{sec:results}
In this section, we present the evaluation results of the scenario defined in Sect.~\ref{sec:experimental} to determine the efficacies of the model in limited data availability across various languages and adaptation strategies. 
Firstly, we evaluate English-trained models on the target languages. Then, we train from scratch in the target language to verify the need for large-scale data training in multilingual settings. We further explore fine-tuning with a single related or unrelated language and with multiple languages while omitting the specific language, thus examining the potential of cross- and intra-linguistic adaptations.  
Intralinguistic strategy, focusing on within-language fine-tuning, demonstrates that even a limited amount of data can meaningfully improve detection accuracy in the target language. Cross-linguistic adaptations, that is the strategy that uses any other language(s) than the target one, offer a promising path for enhancing low-resource language performance through transfer from potential similarities between languages or similar models' interpretations, but the results point out the challenges and limitations, thus leading to using unusual combinations of languages that are not related to each other. 

In Tables \textbf{bold} values indicate the best performance for a specific language (across all models), while \underline{underlined} values highlight the best performance for a specific pre-trained model within each language.

\subsection{Baseline}
\label{ssec:baseline}

\begin{table*}[htbp!]
    \caption{The mean EER scores of baseline models trained with the large English dataset evaluated with the data split procedure described in Section~\ref{sec:experimental}. \textbf{Bold} values indicate the best performance for a specific language. }

    \label{tab: baseline}
    \centering
    \resizebox{0.95\textwidth}{!}{
    \begin{tabular}{c|c|cccccccc}
    \hline
    \multirow{2}{*}{\textbf{Model}} & \multirow{2}{*}{\textbf{Trained with}} & \multicolumn{8}{c}{\textbf{Languages}} \\
    \cline{3-10}
     &  & de & en & es & fr & it & pl & ru & uk \\
    \hline
    \midrule
    {W2V+AASIST} & \multirow{5}{*}{ASVspoof2019} &\textbf{ 2.81 $\pm$ 0.20 }&\textbf{ 1.57 ± 0.14 }& \textbf{2.48 $\pm$ 0.31} & \textbf{0.38 $\pm$ 0.04} & \textbf{0.31 $\pm$ 0.09} & \textbf{0.36 $\pm$ 0.03} & \textbf{15.74 $\pm$ 8.91 } & \textbf{0.60 $\pm$ 0.39} \\ 
    LFCC+AASIST &  & \underline{5.84 $\pm$ 0.44} & \underline{3.74 $\pm$ 1.74} & 24.67 $\pm$ 2.67 & \underline{1.34 $\pm$ 0.34} & \underline{4.53 $\pm$ 0.95} & \underline{1.11 $\pm$ 0.16} & 22.80 $\pm$ 5.62 & \underline{1.39 $\pm$ 1.18} \\ 
    LFCC+MesoNet &  & 6.42 $\pm$ 0.40 & 10.15 $\pm$ 0.76 & \underline{12.24 $\pm$ 2.21} & 2.66 $\pm$ 1.83 & 4.54 $\pm$ 1.15 & 1.76 $\pm$ 0.57 & \underline{17.89 $\pm$ 9.79} & 8.02 $\pm$ 8.36 \\ 
    RawGAT-ST &  & 47.04 $\pm$ 4.73 & 43.58 $\pm$ 2.04 & 41.74 $\pm$ 4.69 & 44.72 $\pm$ 12.60 & 43.40 $\pm$ 13.80 & 41.87 $\pm$ 8.55 & 32.78 $\pm$ 10.13 & 35.46 $\pm$ 6.54 \\ 
    Whisper+AASIST &  & 43.57 $\pm$ 1.48 & 42.76 $\pm$ 1.31 & 42.73 $\pm$ 6.20 & 41.32 $\pm$ 4.93 & 35.52 $\pm$ 13.32 & 42.49 $\pm$ 6.69 & 35.69 $\pm$ 10.14 & 31.95 $\pm$ 6.43 \\ 
    \hline
    \end{tabular}
    }
\end{table*}

The experiment shows that the performance of models trained on the entire ASVspoof2019 LA, comprising 97,168 training samples and 24,293 validation samples, varies very depending on language and model. As shown in Table~\ref{tab: baseline}, W2V+AASIST, pre-trained on an English benchmark dataset, achieves the highest performance across nearly all tested languages, thus substantially outperforming the lowest EERs of other models for each language, presenting superior generalization capabilities across languages.    
While the best LFCC-based model varies across languages, these models trailed behind W2V+AASIST overall and far outperformed alternatives like Whisper+AASIST and RawGAT-ST, demonstrating insufficient cross-lingual generalization capabilities. 

The results show widely differing efficacies of audio DF detection across models and languages. Notably, specific languages revealed more significant challenges for the pre-trained models. In particular, the Russian language exhibits the highest challenge, achieving an EER of $15.74\pm8.91\%$ for W2V+AASIST and $17.89\pm9.79$\%
for LFCC+MesoNet.  Analyzing the best-performing models trained on the English benchmark dataset, it is noteworthy that DFs in some languages are even more detectable than in English, despite being trained only with English samples. 
Specifically, LFCC+AASIST and LFCC+MesoNet detect DFs more effectively in French, Polish, and Ukrainian, while W2V+AASIST shows improved performance in these languages, as well as Italian. 
While Ukrainian's performance can be explained by poorer sample quality compared to other languages, as visible in Table~\ref{tab:edit-distance},  the superior performance of the remaining languages relative to English requires additional analysis in subsequent works.

\subsection{Language training}
\begin{table*}[htbp!]
    \caption{The mean EER scores of trained from scratch with a single language. }
    \label{tab:Trained with_de_en_es_fr_it_pl_ru_uk}
    \centering
    \resizebox{0.95\textwidth}{!}{
    \begin{tabular}{c|c|cccccccc}
    \hline
    \multirow{2}{*}{\textbf{Model}} &\multirow{2}{*}{\textbf{\shortstack{Trained \\ with}}} & \multicolumn{8}{c}{\textbf{Languages}} \\
    \cline{3-10}
 &  & de & en & es & fr & it & pl & ru & uk \\
    \hline
    \midrule
    \multirow{8}{*}{W2V+AASIST}
&de & 13.35 $\pm$ 5.90  & 16.67 $\pm$ 8.19  & 21.22 $\pm$ 5.01  & 10.71 $\pm$ 4.89  & 13.79 $\pm$ 4.45  & 6.80 $\pm$ 4.02   & 25.60 $\pm$ 6.12  & 16.91 $\pm$ 9.58  \\
&en & 20.71 $\pm$ 12.37 & 16.59 $\pm$ 15.43 & 26.44 $\pm$ 17.17 & 21.56 $\pm$ 19.09 & 24.94 $\pm$ 19.09 & 19.88 $\pm$ 18.12 & 36.49 $\pm$ 0.83  & 10.73 $\pm$ 7.80  \\
&es & 25.26 $\pm$ 10.14 & 28.54 $\pm$ 16.25 & 29.10 $\pm$ 18.37 & 20.36 $\pm$ 11.42 & 20.55 $\pm$ 12.28 & 20.34 $\pm$ 11.90 & 29.49 $\pm$ 13.46 & 25.85 $\pm$ 19.35 \\
&fr & 8.60 $\pm$ 3.37   & 13.87 $\pm$ 3.54  & 8.84 $\pm$ 1.64   & 11.28 $\pm$ 1.44  & 5.38 $\pm$ 1.81   & 3.32 $\pm$ 1.73   & 17.25 $\pm$ 6.33  & \textbf{1.33 $\pm$ 0.35}   \\
&it & 28.66 $\pm$ 17.03 & 32.75 $\pm$ 8.83  & 29.22 $\pm$ 11.65 & 25.59 $\pm$ 17.83 & 25.70 $\pm$ 17.27 & 23.71 $\pm$ 17.25 & 34.85 $\pm$ 9.89  & 19.18 $\pm$ 15.42 \\
&pl & 16.10 $\pm$ 7.47  & 32.85 $\pm$ 6.78  & 21.63 $\pm$ 4.38  & 20.14 $\pm$ 0.28  & 13.54 $\pm$ 6.81  & 6.79 $\pm$ 2.42   & 20.33 $\pm$ 1.24  & 21.06 $\pm$ 11.00 \\
&ru & 21.07 $\pm$ 7.69  & 27.57 $\pm$ 0.29  & 18.04 $\pm$ 2.70  & 18.23 $\pm$ 10.17 & 14.20 $\pm$ 6.58  & 13.70 $\pm$ 7.85  & \textbf{12.27 $\pm$ 2.39 } & 9.60 $\pm$ 4.41   \\
&uk & 32.23 $\pm$ 6.17  & 26.99 $\pm$ 3.30  & 28.24 $\pm$ 0.69  & 24.18 $\pm$ 5.15  & 21.84 $\pm$ 5.71  & 24.28 $\pm$ 1.53  & 31.95 $\pm$ 3.26  & 23.26 $\pm$ 6.24  \\
[0.5mm]\hline\noalign{\vskip 0.5mm}
\multirow{8}{*}{LFCC+AASIST}
& de & 7.97 $\pm$ 3.52   & \textbf{4.51 $\pm$ 2.41 }  & 26.63 $\pm$ 12.50 & 5.92 $\pm$ 0.69   & 3.21 $\pm$ 1.24   & 10.43 $\pm$ 8.83  & 19.82 $\pm$ 18.86 & 2.35 $\pm$ 0.79   \\
& en & 7.71 $\pm$ 1.66   & \underline{4.93 $\pm$ 0.11}   & 16.83 $\pm$ 4.15  & 5.11 $\pm$ 0.09   & \underline{1.43 $\pm$ 0.31}   & 2.67 $\pm$ 0.19   & \underline{14.18 $\pm$ 13.75} & 7.43 $\pm$ 7.78   \\
& es & 9.23 $\pm$ 1.16   & 4.98 $\pm$ 0.48   & 20.51 $\pm$ 20.83 & \underline{4.36 $\pm$ 0.15}   & \textbf{1.32 $\pm$ 0.85}   & 1.78 $\pm$ 0.65   & 25.49 $\pm$ 26.04 & 3.31 $\pm$ 0.33   \\
& fr & 9.16 $\pm$ 1.75   & 6.74 $\pm$ 0.33   & \underline{7.00 $\pm$ 2.42}   & 5.13 $\pm$ 0.08   & 3.77 $\pm$ 0.62   & \textbf{1.11 $\pm$ 0.13}   & 16.78 $\pm$ 17.49 & 2.69 $\pm$ 2.84   \\
& it & 8.21 $\pm$ 1.04   & 7.27 $\pm$ 0.29   & \textbf{5.69 $\pm$ 4.67}   & 6.06 $\pm$ 0.08   & 4.30 $\pm$ 0.25   & 2.26 $\pm$ 0.28   & 17.82 $\pm$ 18.78 & 2.86 $\pm$ 0.91   \\
 & pl & 10.09 $\pm$ 0.59  & 7.77 $\pm$ 0.99   & 16.05 $\pm$ 5.61  & 6.14 $\pm$ 0.15   & 2.02 $\pm$ 0.09   & \underline{1.31 $\pm$ 0.06}   & 16.88 $\pm$ 16.12 & 5.00 $\pm$ 5.21   \\
& ru & 15.48 $\pm$ 8.91  & 16.26 $\pm$ 16.09 & 34.01 $\pm$ 17.64 & 9.16 $\pm$ 9.55   & 9.09 $\pm$ 8.61   & 7.55 $\pm$ 7.69   & 24.76 $\pm$ 19.50 & 7.63 $\pm$ 1.03   \\
& uk & 10.46 $\pm$ 1.41  & 7.12 $\pm$ 1.12   & 12.24 $\pm$ 1.94  & 4.59 $\pm$ 0.52   & 1.89 $\pm$ 0.23   & 4.44 $\pm$ 0.21   & 18.27 $\pm$ 15.66 & \underline{1.49 $\pm$ 0.21}   \\
[0.5mm]\hline\noalign{\vskip 0.5mm}
\multirow{8}{*}{LFCC+MesoNet}
&de & 13.41 $\pm$ 3.64  & 8.84 $\pm$ 6.34   & 24.12 $\pm$ 0.31  & 5.15 $\pm$ 4.76   & 9.01 $\pm$ 3.93   & 13.10 $\pm$ 12.53 & 30.48 $\pm$ 12.81 & 6.74 $\pm$ 7.03   \\
& en & 7.78 $\pm$ 0.48   & 6.80 $\pm$ 0.38   & 13.36 $\pm$ 5.47  & 7.16 $\pm$ 0.02   & 1.99 $\pm$ 0.40   & 3.73 $\pm$ 0.29   & 17.82 $\pm$ 14.55 & 3.21 $\pm$ 0.23   \\
&es & 10.95 $\pm$ 0.24  & 10.66 $\pm$ 2.85  & 13.91 $\pm$ 2.27  & \textbf{2.50 $\pm$ 2.04}   & 7.09 $\pm$ 3.05   & 2.33 $\pm$ 0.56   & 22.63 $\pm$ 18.00 & 4.90 $\pm$ 5.05   \\
&fr & \underline{7.12 $\pm$ 1.56}   & 13.06 $\pm$ 0.42  & 16.19 $\pm$ 9.08  & 6.20 $\pm$ 0.13   & 2.31 $\pm$ 0.38   & 5.92 $\pm$ 0.46   & 21.90 $\pm$ 9.65  & 7.67 $\pm$ 0.70   \\
&it & 8.91 $\pm$ 1.09   & 5.39 $\pm$ 2.46   & 16.52 $\pm$ 2.80  & 7.50 $\pm$ 0.99   & 2.82 $\pm$ 1.66   & 2.84 $\pm$ 1.86   & 15.65 $\pm$ 14.79 & 2.96 $\pm$ 3.08   \\
&pl & \textbf{6.81 $\pm$ 1.05}   & 31.22 $\pm$ 1.64  & 17.75 $\pm$ 7.01  & 10.38 $\pm$ 0.22  & 2.66 $\pm$ 0.64   & 4.74 $\pm$ 0.09   & 18.91 $\pm$ 10.69 & 3.85 $\pm$ 4.06   \\
&ru & 17.52 $\pm$ 12.41 & 23.43 $\pm$ 17.51 & 27.47 $\pm$ 11.93 & 12.96 $\pm$ 12.97 & 17.41 $\pm$ 15.59 & 14.03 $\pm$ 13.46 & 25.70 $\pm$ 10.51 & 4.08 $\pm$ 0.48   \\
&uk & 9.59 $\pm$ 2.42   & 6.57 $\pm$ 2.28   & 23.14 $\pm$ 11.66 & 6.61 $\pm$ 0.15   & 4.28 $\pm$ 2.31   & 1.55 $\pm$ 0.47   & 24.64 $\pm$ 2.87  & 2.83 $\pm$ 2.98   \\
[0.5mm]\hline\noalign{\vskip 0.5mm}
\bottomrule
\end{tabular}
}
\end{table*}

To investigate the effectiveness of limited training data, we train models from scratch \textbf{with a single language} following the dataset partitioning described in~Section~\ref{sec:experimental}. The aim is to assess whether even a small number of samples for a particular language and training the model could be an alternative to long training on a large dataset. The SSL-based front-end is re-initialized as in previous experiments. 

The results presented in Table~\ref{tab:Trained with_de_en_es_fr_it_pl_ru_uk} indicate that models trained from scratch generally perform poorly, especially compared to the pre-trained models, which overreach in every scenario. However, LFCC+AASIST achieves the best results relatively across all models. Nevertheless, only Russian achieves slightly better performance than pre-trained models with the English benchmark.  On the other hand, for the remaining languages, we observe a significant drop in detection efficacy. Therefore, we cannot replace large-scale pretraining with small, targeted data. Since W2V+AASIST and LFCC-based models significantly outperform RawGAT-ST and Whisper-AASIST in our experiments, we report only the results for the former models and place the latter in Appendix~\ref{app:additional results}.

These results confirm that having a large, language-independent amount of data enhances detection more than small, language-specific datasets, thus highlighting and reinforcing the value of extensive, even language-independent, data for DF detection.

\begin{table*}[htbp!]
    \caption{The mean EER scores of fine-tuned \textbf{with} a specific language with the data split procedure described in Section~\ref{sec:experimental}.}
    \label{tab: Fine-tuned with}
    \centering
    \resizebox{0.95\textwidth}{!}{
    \begin{tabular}{c|c|cccccccc}
    \hline
    \multirow{2}{*}{\textbf{Model}} & \multirow{2}{*}{\textbf{\shortstack{Fine-tuned \\ with}}} & \multicolumn{8}{c}{\textbf{Languages}} \\
    \cline{3-10}
 &  & de & en & es & fr & it & pl & ru & uk \\
    \hline
    \midrule
    \multirow{8}{*}{W2V+AASIST}
& de & \underline{\textbf{1.65 $\pm$ 0.71}} & 4.16 $\pm$ 2.07 & 5.59 $\pm$ 1.56 & 0.70 $\pm$ 0.28 & 1.49 $\pm$ 0.18 & 0.84 $\pm$ 0.05 & 17.80 $\pm$ 9.24 & 0.35 $\pm$ 0.13 \\ 
& en & 4.07 $\pm$ 1.02 & \underline{\textbf{0.63 $\pm$ 0.10}} & 2.29 $\pm$ 0.74 & 0.26 $\pm$ 0.10 & 0.49 $\pm$ 0.28 & 0.24 $\pm$ 0.04 & 22.97 $\pm$ 5.57 & 0.13 $\pm$ 0.10 \\ 
& es & 4.32 $\pm$ 0.65 & 2.80 $\pm$ 1.92 & \underline{\textbf{1.10 $\pm$ 0.34 }}& 0.22 $\pm$ 0.08 & 0.23 $\pm$ 0.04 & 0.29 $\pm$ 0.11 & 16.19 $\pm$ 10.34 & 0.16 $\pm$ 0.05 \\ 
& fr & 3.01 $\pm$ 0.45 & 3.15 $\pm$ 2.10 & 4.62 $\pm$ 0.65 &\underline{\textbf{ 0.11 $\pm$ 0.06}} & 0.33 $\pm$ 0.07 & 0.24 $\pm$ 0.04 & 20.90 $\pm$ 7.22 & \underline{\textbf{0.06 $\pm$ 0.06}} \\ 
& it & 3.81 $\pm$ 0.58 & 3.74 $\pm$ 2.63 & 3.49 $\pm$ 0.19 & 0.56 $\pm$ 0.44 & \underline{\textbf{0.22 $\pm$ 0.14}} & 0.28 $\pm$ 0.08 & 19.62 $\pm$ 9.27 & 0.11 $\pm$ 0.03 \\ 
& pl & 2.48 $\pm$ 0.57 & 4.58 $\pm$ 3.87 & 3.36 $\pm$ 0.50 & 0.42 $\pm$ 0.19 & 0.42 $\pm$ 0.21 & \underline{\textbf{0.19 $\pm$ 0.08}} & 18.58 $\pm$ 8.42 & 0.20 $\pm$ 0.08 \\ 
& ru & 3.73 $\pm$ 1.28 & 4.76 $\pm$ 2.32 & 1.69 $\pm$ 0.17 & 0.90 $\pm$ 0.24 & 0.91 $\pm$ 0.10 & 0.50 $\pm$ 0.26 & \underline{\textbf{3.59 $\pm$ 3.03}} & 0.76 $\pm$ 0.08 \\ 
& uk & 2.52 $\pm$ 0.57 & 2.50 $\pm$ 1.72 & 3.01 $\pm$ 0.26 & 0.26 $\pm$ 0.06 & 0.28 $\pm$ 0.06 & 0.36 $\pm$ 0.18 & 18.08 $\pm$ 9.92 & 0.42 $\pm$ 0.35 \\ 
[0.5mm]\hline\noalign{\vskip 0.5mm}
\multirow{8}{*}{LFCC+AASIST}
& de & \underline{5.52 $\pm$ 0.81} & 4.04 $\pm$ 2.89 & 20.05 $\pm$ 1.57 & 1.15 $\pm$ 0.79 & 3.46 $\pm$ 0.99 & 2.36 $\pm$ 1.72 & 24.00 $\pm$ 5.94 & 1.45 $\pm$ 1.41 \\ 
& en & 6.89 $\pm$ 1.70 & 2.13 $\pm$ 1.05 & 20.46 $\pm$ 3.49 & 0.34 $\pm$ 0.09 & 2.56 $\pm$ 0.26 & 0.88 $\pm$ 0.15 & 24.47 $\pm$ 5.56 & 1.20 $\pm$ 1.05 \\ 
& es & 8.07 $\pm$ 2.13 & 3.71 $\pm$ 0.10 & \underline{12.51 $\pm$ 1.63} & 0.48 $\pm$ 0.35 & 3.44 $\pm$ 0.17 & 1.07 $\pm$ 0.11 & 16.70 $\pm$ 10.07 & 0.98 $\pm$ 0.73 \\ 
& fr & 7.39 $\pm$ 0.64 & 1.44 $\pm$ 0.62 & 19.93 $\pm$ 6.28 & \underline{0.22 $\pm$ 0.19} & 2.32 $\pm$ 0.35 & \underline{0.44 $\pm$ 0.11} & 25.80 $\pm$ 5.49 & 0.65 $\pm$ 0.61 \\ 
& it & 7.15 $\pm$ 0.56 & \underline{1.43 $\pm$ 0.24} & 13.43 $\pm$ 1.14 & 0.41 $\pm$ 0.07 & \underline{1.22 $\pm$ 0.12} & 0.52 $\pm$ 0.06 & 18.90 $\pm$ 10.63 & 0.77 $\pm$ 0.78 \\ 
& pl & 8.28 $\pm$ 1.40 & 3.77 $\pm$ 2.63 & 20.55 $\pm$ 6.27 & 0.49 $\pm$ 0.17 & 2.41 $\pm$ 0.56 & 0.69 $\pm$ 0.16 & 23.84 $\pm$ 5.73 & 1.37 $\pm$ 1.26 \\ 
& ru & 7.70 $\pm$ 1.85 & 4.84 $\pm$ 3.03 & 22.47 $\pm$ 9.20 & 2.80 $\pm$ 2.19 & 7.22 $\pm$ 4.17 & 2.81 $\pm$ 1.90 & \underline{14.07 $\pm$ 9.57} & 1.17 $\pm$ 0.96 \\ 
& uk & 7.23 $\pm$ 2.01 & 5.31 $\pm$ 4.17 & 21.06 $\pm$ 2.82 & 0.55 $\pm$ 0.08 & 3.11 $\pm$ 0.80 & 0.74 $\pm$ 0.09 & 24.48 $\pm$ 4.27 & \underline{0.36 $\pm$ 0.23} \\ 
[0.5mm]\hline\noalign{\vskip 0.5mm}
\multirow{8}{*}{LFCC+MesoNet}
& de & 17.55 $\pm$ 0.50 & 17.65 $\pm$ 3.89 & 18.75 $\pm$ 5.41 & 11.33 $\pm$ 3.75 & 13.61 $\pm$ 0.39 & 9.15 $\pm$ 6.13 & 21.94 $\pm$ 11.46 & 23.04 $\pm$ 23.91 \\ 
& en & 6.29 $\pm$ 0.43 & 6.55 $\pm$ 3.51 & 15.80 $\pm$ 4.84 & 0.69 $\pm$ 0.11 & 2.92 $\pm$ 0.80 & 0.56 $\pm$ 0.32 & 18.64 $\pm$ 8.79 & 6.98 $\pm$ 7.32 \\ 
& es & 9.58 $\pm$ 3.98 & 11.15 $\pm$ 6.97 & 18.85 $\pm$ 2.19 & 3.51 $\pm$ 0.38 & 9.65 $\pm$ 0.85 & 4.15 $\pm$ 2.58 & 22.96 $\pm$ 5.60 & 20.13 $\pm$ 21.18 \\ 
& fr & 7.55 $\pm$ 0.53 & 12.51 $\pm$ 0.92 & 13.37 $\pm$ 2.01 & 2.99 $\pm$ 2.00 & 6.01 $\pm$ 1.69 & 1.57 $\pm$ 0.55 & 18.67 $\pm$ 9.34 & 8.15 $\pm$ 8.51 \\ 
& it & 7.53 $\pm$ 1.46 & 11.42 $\pm$ 2.51 & 14.54 $\pm$ 3.94 & 2.34 $\pm$ 0.96 & 5.63 $\pm$ 0.09 & 2.18 $\pm$ 1.37 & 18.31 $\pm$ 10.16 & 11.25 $\pm$ 11.80 \\ 
& pl & 8.06 $\pm$ 3.76 & 11.47 $\pm$ 6.94 & 15.78 $\pm$ 4.53 & 2.12 $\pm$ 0.65 & 5.53 $\pm$ 1.90 & 3.37 $\pm$ 3.13 & 19.33 $\pm$ 10.35 & 17.98 $\pm$ 18.92 \\ 
& ru & 20.63 $\pm$ 12.80 & 23.28 $\pm$ 10.78 & 21.04 $\pm$ 0.49 & 25.35 $\pm$ 25.02 & 21.34 $\pm$ 17.05 & 19.00 $\pm$ 15.10 & 34.20 $\pm$ 4.33 & 9.06 $\pm$ 0.49 \\ 
& uk & 7.82 $\pm$ 2.92 & 9.91 $\pm$ 5.12 & 15.45 $\pm$ 5.74 & 1.58 $\pm$ 0.37 & 5.50 $\pm$ 2.08 & 2.40 $\pm$ 2.12 & 18.80 $\pm$ 10.98 & 8.42 $\pm$ 8.84 \\ 
[0.5mm]
\hline

    \bottomrule
    \end{tabular}
    }
    \end{table*}

\begin{table*}[hbp]
    \caption{The mean EER scores of  fine-tuned \textbf{without} a single language.}
    \label{tab: Fine-tuned-without}
    \centering
    \resizebox{0.95\textwidth}{!}{
    \begin{tabular}{c|c|cccccccc}
    \hline
    \multirow{2}{*}{\textbf{Model}} & \multirow{2}{*}{\textbf{\shortstack{Fine-tuned \\ without}}} & \multicolumn{8}{c}{\textbf{Languages}} \\
    \cline{3
    -10}
 &  & de & en & es & fr & it & pl & ru & uk \\
    \hline
    \midrule
    \multirow{8}{*}{W2V+AASIST}
& de & 3.61 $\pm$ 0.25 & \underline{\textbf{0.45 $\pm$ 0.09}} & \underline{\textbf{0.67 $\pm$ 0.53}} & \underline{\textbf{0.04 $\pm$ 0.05}} & \underline{\textbf{0.12 $\pm$ 0.03}} & \underline{\textbf{0.06 $\pm$ 0.03}} & 10.81 $\pm$ 11.18 &\underline{\textbf{ 0.02 $\pm$ 0.03}} \\ 
& en & \underline{\textbf{1.09 $\pm$ 0.50}} & 1.94 $\pm$ 1.12 & 1.18 $\pm$ 0.82 & 0.15 $\pm$ 0.12 & 0.41 $\pm$ 0.11 & 0.08 $\pm$ 0.03 & 7.95 $\pm$ 8.05 & 0.08 $\pm$ 0.03 \\ 
& es & 1.36 $\pm$ 0.29 & 0.52 $\pm$ 0.05 & 2.60 $\pm$ 0.68 & 0.07 $\pm$ 0.04 & 0.51 $\pm$ 0.14 & 0.13 $\pm$ 0.03 & 12.50 $\pm$ 12.78 & 0.06 $\pm$ 0.06 \\ 
& fr & 1.86 $\pm$ 0.21 & 0.76 $\pm$ 0.05 & 1.37 $\pm$ 0.79 & 0.47 $\pm$ 0.45 & 0.52 $\pm$ 0.11 & 0.16 $\pm$ 0.05 & 10.92 $\pm$ 10.95 & 0.08 $\pm$ 0.03 \\ 
& it & 1.48 $\pm$ 0.21 & 0.70 $\pm$ 0.08 & 1.74 $\pm$ 1.44 & 0.11 $\pm$ 0.12 & 0.66 $\pm$ 0.06 & 0.11 $\pm$ 0.05 & 10.60 $\pm$ 10.72 & 0.06 $\pm$ 0.06 \\ 
& pl & 1.43 $\pm$ 0.46 & 0.55 $\pm$ 0.07 & 1.26 $\pm$ 1.00 & 0.05 $\pm$ 0.02 & 0.26 $\pm$ 0.12 & 0.17 $\pm$ 0.12 & 11.61 $\pm$ 11.92 & 0.06 $\pm$ 0.06 \\ 
& ru & 1.51 $\pm$ 0.30 & 0.45 $\pm$ 0.24 & 0.87 $\pm$ 0.54 & 0.13 $\pm$ 0.14 & 0.21 $\pm$ 0.14 & 0.12 $\pm$ 0.04 & 22.62 $\pm$ 8.35 & 0.08 $\pm$ 0.03 \\ 
& uk & 1.56 $\pm$ 0.60 & 0.62 $\pm$ 0.24 & 1.47 $\pm$ 1.20 & 0.18 $\pm$ 0.14 & 0.38 $\pm$ 0.25 & 0.22 $\pm$ 0.07 & \underline{\textbf{7.50 $\pm$ 7.59}} & 0.03 $\pm$ 0.03 \\ 
[0.5mm]\hline\noalign{\vskip 0.5mm}
\multirow{8}{*}{LFCC+AASIST}
& de & 7.99 $\pm$ 1.57 & \underline{2.05 $\pm$ 0.85} & \underline{7.95 $\pm$ 3.22} & \underline{0.15 $\pm$ 0.05} & \underline{1.13 $\pm$ 0.19} & 0.46 $\pm$ 0.06 & 18.53 $\pm$ 18.58 & \underline{0.17 $\pm$ 0.19} \\ 
& en & 4.51 $\pm$ 0.30 & 3.84 $\pm$ 1.18 & 8.79 $\pm$ 3.50 & 0.35 $\pm$ 0.10 & 1.60 $\pm$ 0.18 & 0.50 $\pm$ 0.28 & 15.40 $\pm$ 15.05 & 0.27 $\pm$ 0.29 \\ 
& es & 4.58 $\pm$ 0.47 & 3.55 $\pm$ 0.22 & 16.21 $\pm$ 5.88 & 0.25 $\pm$ 0.13 & 1.72 $\pm$ 0.10 & 0.51 $\pm$ 0.26 & \underline{14.53 $\pm$ 13.94} & 0.25 $\pm$ 0.26 \\ 
& fr & 4.84 $\pm$ 0.23 & 3.47 $\pm$ 1.79 & 9.04 $\pm$ 5.84 & 0.51 $\pm$ 0.10 & 1.67 $\pm$ 0.65 & 0.82 $\pm$ 0.61 & 15.23 $\pm$ 15.23 & 0.51 $\pm$ 0.53 \\ 
& it & 5.37 $\pm$ 1.19 & 3.94 $\pm$ 1.11 & 9.19 $\pm$ 4.11 & 0.37 $\pm$ 0.17 & 2.08 $\pm$ 0.37 & 0.84 $\pm$ 0.56 & 15.48 $\pm$ 15.28 & 0.32 $\pm$ 0.27 \\ 
& pl & 5.18 $\pm$ 0.93 & 3.19 $\pm$ 0.87 & 8.77 $\pm$ 3.80 & 0.38 $\pm$ 0.08 & 1.43 $\pm$ 0.45 & 0.52 $\pm$ 0.26 & 15.44 $\pm$ 15.12 & 0.31 $\pm$ 0.33 \\ 
& ru &\underline{3.58 $\pm$ 0.75} & 2.69 $\pm$ 2.23 & 8.01 $\pm$ 7.63 & 0.31 $\pm$ 0.23 & 1.17 $\pm$ 0.85 & \underline{0.42 $\pm$ 0.32} & 15.44 $\pm$ 15.56 & 0.47 $\pm$ 0.49 \\ 
& uk & 6.45 $\pm$ 1.74 & 3.86 $\pm$ 1.25 & 10.20 $\pm$ 3.40 & 0.48 $\pm$ 0.03 & 1.82 $\pm$ 0.34 & 0.66 $\pm$ 0.22 & 14.93 $\pm$ 14.59 & 0.53 $\pm$ 0.53 \\ 
[0.5mm]\hline\noalign{\vskip 0.5mm}
\multirow{8}{*}{LFCC+MesoNet}
& de & 5.51 $\pm$ 0.35 & 7.46 $\pm$ 2.24 & 10.54 $\pm$ 1.82 & 1.95 $\pm$ 1.33 & 4.05 $\pm$ 1.62 & 0.88 $\pm$ 0.29 & 22.23 $\pm$ 19.57 & 2.46 $\pm$ 2.56 \\ 
& en & 11.62 $\pm$ 2.64 & 13.48 $\pm$ 6.20 & 16.30 $\pm$ 7.07 & 6.73 $\pm$ 1.20 & 9.92 $\pm$ 0.49 & 4.39 $\pm$ 3.26 & 23.93 $\pm$ 17.59 & 14.68 $\pm$ 15.39 \\ 
& es & 11.23 $\pm$ 1.78 & 12.68 $\pm$ 5.89 & 15.63 $\pm$ 6.20 & 6.27 $\pm$ 2.07 & 9.14 $\pm$ 0.86 & 3.97 $\pm$ 2.95 & 25.30 $\pm$ 18.83 & 9.83 $\pm$ 10.28 \\ 
& fr & 13.00 $\pm$ 5.29 & 14.14 $\pm$ 7.70 & 16.64 $\pm$ 8.66 & 7.87 $\pm$ 2.10 & 11.08 $\pm$ 3.44 & 7.08 $\pm$ 6.46 & 25.32 $\pm$ 20.48 & 15.77 $\pm$ 16.58 \\ 
& it & 12.76 $\pm$ 2.44 & 13.14 $\pm$ 6.76 & 17.21 $\pm$ 6.90 & 8.09 $\pm$ 1.42 & 10.61 $\pm$ 1.25 & 5.45 $\pm$ 4.05 & 26.15 $\pm$ 19.00 & 12.19 $\pm$ 12.77 \\ 
& pl & 12.75 $\pm$ 3.19 & 13.05 $\pm$ 6.96 & 17.42 $\pm$ 6.58 & 7.85 $\pm$ 0.67 & 10.73 $\pm$ 1.51 & 5.37 $\pm$ 4.12 & 26.05 $\pm$ 19.30 & 11.88 $\pm$ 12.44 \\ 
& ru & 7.52 $\pm$ 2.64 & 8.50 $\pm$ 7.63 & 16.49 $\pm$ 2.14 & 1.48 $\pm$ 1.33 & 4.79 $\pm$ 2.84 & 1.97 $\pm$ 1.81 & 20.33 $\pm$ 12.74 & 8.48 $\pm$ 8.94 \\ 
& uk & 12.36 $\pm$ 5.56 & 13.99 $\pm$ 7.55 & 17.51 $\pm$ 7.96 & 7.04 $\pm$ 2.07 & 10.21 $\pm$ 4.01 & 6.68 $\pm$ 6.11 & 24.44 $\pm$ 20.52 & 14.75 $\pm$ 15.46 \\ 
\hline
    \bottomrule
    \end{tabular}
    }
    \end{table*}
\subsection{Language fine-tuning}
\label{ssec:ft}

Fine-tuning pre-trained models with English benchmark data allows us to assess the need for fine-tuning to improve models' detectability and potential cross-language generalization capabilities. Fine-tuning and further evaluation follow the data split described in Section~\ref{sec:experimental}. 

We first assess whether fine-tuning with single-language data enhances audio DF detection by adding linguistic context, improving performance over English pre-trained models. Based on the results in Table~\ref{tab: Fine-tuned with}, we can distinguish two trends. In the first one, intra-linguistic adaptation is more efficient and thus reduces the EER compared to pre-trained and cross-adaptation models. This group includes better-performing models: LFCC+AASIST and W2V+AASIST. 
On the other hand, the second group shows a trend that fine-tuning with a specific language is most effective. Specifically, RawGAT-ST, fine-tuned with Polish, and Whisper+AASIST, as well as LFCC+MesoNet, fine-tuned with English, achieve the lowest EER across most languages for this specific model. 

A deeper analysis reveals that W2V+AASIST consistently outperforms other architectures, achieving the lowest EER for most languages, confirming the effectiveness of intralinguistic adaptation. Cross-lingual adaptation remains crucial for multilingual models, with W2V+AASIST showing competitive results when the pre-trained model performs well. However, their performance remains comparable to or worse than the baseline for languages like Russian and some Germanic languages. Results for LFCC+AASIST indicate similar trends with the best results on the diagonal, thus indicating that intra-language adaptations are more effective. Notably, cross-linguistic fine-tuning of the LFCC+AASIST is effective in two scenarios: improving a well-performing pre-trained model (e.g., in the case of French or Polish) or fine-tuning with Italian, which indicates overperforming other even intralinguistic adaptations.

The further investigation focuses on two key aspects of linguistic adaptability in DF detection: the impact of removing a single language from fine-tuning and the trade-off between language-specific and multilingual training. We fine-tune pre-trained models on \textbf{all multilingual data except one language} at a time, following the data split detailed in Section~\ref{sec:experimental}. During these experiments, we assess whether combining multiple languages for training might provide comparable results, especially in the context of cross-language adaptations. 
As shown in Table~\ref{tab: Fine-tuned-without}, fine-tuning with limited language-specific data generally outperforms a larger multilingual dataset, excluding the target language, with Ukrainian as the only exception. Intriguingly, our analysis reveals that if one language is excluded to optimize the system,  German emerges as the most suitable candidate for removal from the training set. This is only relevant for scenarios where German language detection is not a requirement, as its exclusion demonstrated a positive effect on the overall system performance for almost all other languages. 

Our results suggest maximizing language coverage in training data whenever feasible. For language-specific deployments, focus on using relevant language data, as even limited amounts demonstrate more usefulness than larger, more linguistically diverse datasets that lack the target language.

\noindent\textbf{Answer to RQ1}: 
\noindent The detection of audio DF shows significant variability across languages. The evaluation of English pre-trained models without additional linguistic adaptation reveals that models like W2V+AASIST show cross-lingual generalization capabilities, yet their performance varied significantly across languages. The efficacy of some languages is even better than that of the only language \textit{known to the model}, English. On the other hand, the Russian language reveals limitations in this approach, thus highlighting the critical necessity for adapting even the best-performing models to other languages.

\noindent\textbf{Answer to RQ2}: 
Evaluation using English pre-trained models without additional adaptation reveals notable differences in performance when applied to other languages.
Languages such as French, Polish, Italian, and Ukrainian show better detection performance than English, even though the models are trained exclusively on English datasets. Conversely, Russian or Spanish seems more challenging, with pre-trained models yielding higher EERs. Nevertheless, for these languages, the intralinguistic adaptations significantly improve audio DF detection.

\noindent\textbf{Answer to RQ3}:
\noindent  The experiments indicate that intralinguistic adaptation is the most effective targeted strategy for improving audio DF detection within specific languages. Our results show that the optimal strategy combines English-pre-trained data with multilingual fine-tuning that includes the target language. Even limited target language data significantly improves detection accuracy in the best-performing models. However, fine-tuning with non-target languages shows varying results, and using either unrelated languages or multilingual datasets without the target language fails to improve performance. Furthermore, our results indicate that fine-tuning on a small corpus of target language data consistently outperforms augmenting the training set with a broader but non-target multilingual mix. In practice, improving detection requires a small set of samples in the language you are interested in, rather than a larger set containing non-target language samples.

\section{Conclusion}
\label{sec:conclusion}
This research aims to introduce a benchmark and answer the question of the effectiveness of pre-trained and adapted audio DF detection models in multilingual settings. Up to now, the influence of the utterances' language on audio DF detection has not been examined unbiasedly. 
The experiments, especially for best-performed W2V+AASIST and LFCC+AASIST, demonstrate a key finding across most of the tested languages: a small amount of language-specific data often yields greater improvements in detection than much more multilingual data, but without the target language.
Although our study is limited to eight languages, samples were generated using a selective range of methods, which affected the confidence of the results. Despite these limitations, we believe that it is an important starting point for tackling the problems and challenges posed by multilingual DFs. Future research should explore more languages and language families, as well as generators and models, to better understand cross-lingual fine-tuning and transfer effects. Additionally, a comparison between tonal and non-tonal languages would be interesting, as well as exploring emotionally varied samples in various languages. 

\section{Acknowledgment}
We gratefully acknowledge Polish high-performance computing infrastructure PLGrid (HPC Centers: WCSS, ACK Cyfronet AGH) for providing computer facilities and support within computational grant no. PLG/2024/017432.

\bibliographystyle{IEEEtran}
\bibliography{mybib}

\appendix
\section{Additional results}
The following tables (Table~\ref{tab:Trained-app-with_de_en_es_fr_it_pl_ru_uk} - Table~\ref{tab: Fine-tuned-without app}) present the experimental results for RawGAT-ST and Whisper-AASIST models.  These models consistently underperformed relative to W2V+AASIST and LFCC-based approaches. 
\label{app:additional results}
\begin{table*}[ht]
    \caption{The mean EER scores of trained from scratch with a single language. }
    \label{tab:Trained-app-with_de_en_es_fr_it_pl_ru_uk}
    \centering
    \resizebox{\textwidth}{!}{
    \begin{tabular}{c|c|cccccccc}
    \hline
    \multirow{2}{*}{\textbf{Model}} &\multirow{2}{*}{\textbf{\shortstack{Trained \\ with}}} & \multicolumn{8}{c}{\textbf{Languages}} \\
    \cline{3-10}
 &  & de & en & es & fr & it & pl & ru & uk \\
    \hline
    \midrule
\multirow{8}{*}{RawGAT-ST}
& de & 50.96 $\pm$ 4.94 & 53.41 $\pm$ 11.67 & 60.26 $\pm$ 0.63 & 50.33 $\pm$ 2.76 & 50.37 $\pm$ 8.11 & 46.44 $\pm$ 6.24 & 54.03 $\pm$ 8.84 & 52.84 $\pm$ 6.01 \\ 
& en & 44.99 $\pm$ 1.58 & 37.74 $\pm$ 9.38 & 49.87 $\pm$ 4.15 & 54.00 $\pm$ 2.05 & 44.29 $\pm$ 2.06 & 50.58 $\pm$ 6.72 & 51.70 $\pm$ 9.92 & 53.37 $\pm$ 10.28 \\ 
& es & 51.62 $\pm$ 10.80 & 74.85 $\pm$ 2.36 & 59.63 $\pm$ 0.46 & 55.15 $\pm$ 10.88 & 56.11 $\pm$ 14.28 & 61.83 $\pm$ 21.78 & 57.70 $\pm$ 1.57 & 52.04 $\pm$ 17.29 \\ 
& fr & 49.69 $\pm$ 8.56 & 52.09 $\pm$ 4.88 & 58.38 $\pm$ 0.59 & 48.07 $\pm$ 4.47 & 50.17 $\pm$ 6.10 & 54.79 $\pm$ 11.27 & 58.34 $\pm$ 4.26 & 47.60 $\pm$ 1.73 \\ 
& it & 48.59 $\pm$ 11.34 & 59.06 $\pm$ 8.02 & 56.33 $\pm$ 0.36 & 55.26 $\pm$ 8.43 & 60.68 $\pm$ 8.01 & 57.41 $\pm$ 13.93 & 49.38 $\pm$ 7.12 & 40.91 $\pm$ 5.44 \\ 
& pl & 54.51 $\pm$ 2.00 & 45.62 $\pm$ 2.60 & 52.42 $\pm$ 8.18 & 48.13 $\pm$ 3.55 & 54.56 $\pm$ 3.35 & 34.80 $\pm$ 10.42 & 53.27 $\pm$ 7.74 & 67.03 $\pm$ 2.62 \\ 
& ru & 36.54 $\pm$ 9.92 & 57.12 $\pm$ 0.51 & 53.45 $\pm$ 3.46 & 45.12 $\pm$ 0.34 & 39.79 $\pm$ 15.07 & 41.99 $\pm$ 2.34 & 36.72 $\pm$ 0.72 & 22.63 $\pm$ 9.60 \\ 
& uk & 58.55 $\pm$ 5.02 & 68.47 $\pm$ 3.28 & 55.68 $\pm$ 3.38 & 51.25 $\pm$ 12.93 & 52.60 $\pm$ 12.05 & 64.26 $\pm$ 15.54 & 45.23 $\pm$ 2.73 & 39.66 $\pm$ 2.64 \\ 
[0.5mm]\hline\noalign{\vskip 0.5mm}
\multirow{8}{*}{Whisper+AASIST}
& de & 53.55 $\pm$ 0.23 & 45.75 $\pm$ 5.32 & 56.27 $\pm$ 1.93 & 46.28 $\pm$ 6.83 & 51.21 $\pm$ 1.82 & 54.00 $\pm$ 4.17 & 51.35 $\pm$ 1.90 & 45.34 $\pm$ 12.18 \\ 
& en & 46.20 $\pm$ 5.11 & 34.78 $\pm$ 1.54 & 48.82 $\pm$ 5.32 & 44.20 $\pm$ 1.76 & 45.52 $\pm$ 6.28 & 50.87 $\pm$ 12.14 & 48.31 $\pm$ 6.86 & 49.81 $\pm$ 9.56 \\ 
& es & 55.39 $\pm$ 1.11 & 49.36 $\pm$ 5.64 & 54.89 $\pm$ 0.63 & 48.19 $\pm$ 6.19 & 57.61 $\pm$ 0.28 & 58.28 $\pm$ 2.00 & 54.17 $\pm$ 2.36 & 49.46 $\pm$ 7.90 \\ 
& fr & 48.13 $\pm$ 4.79 & 43.10 $\pm$ 1.18 & 50.11 $\pm$ 3.77 & 40.00 $\pm$ 0.21 & 48.06 $\pm$ 5.84 & 52.90 $\pm$ 8.07 & 44.38 $\pm$ 6.18 & 41.78 $\pm$ 14.32 \\ 
& it & 50.16 $\pm$ 0.31 & 45.04 $\pm$ 3.49 & 52.52 $\pm$ 1.88 & 44.22 $\pm$ 5.14 & 51.32 $\pm$ 0.62 & 54.98 $\pm$ 5.73 & 48.57 $\pm$ 2.87 & 42.45 $\pm$ 14.66 \\ 
& pl & 54.58 $\pm$ 0.96 & 42.62 $\pm$ 5.01 & 57.16 $\pm$ 2.46 & 47.76 $\pm$ 6.37 & 52.42 $\pm$ 3.28 & 53.71 $\pm$ 2.42 & 52.61 $\pm$ 0.33 & 48.32 $\pm$ 15.14 \\ 
& ru & 51.77 $\pm$ 1.68 & 46.05 $\pm$ 3.02 & 52.03 $\pm$ 1.86 & 42.96 $\pm$ 6.62 & 51.22 $\pm$ 2.65 & 53.27 $\pm$ 3.72 & 43.02 $\pm$ 0.90 & 39.01 $\pm$ 11.99 \\ 
& uk & 48.64 $\pm$ 4.94 & 49.39 $\pm$ 3.97 & 49.34 $\pm$ 6.08 & 46.68 $\pm$ 6.95 & 43.53 $\pm$ 13.28 & 44.05 $\pm$ 10.94 & 42.37 $\pm$ 8.91 & 34.30 $\pm$ 2.03 \\ 
[0.5mm]\hline\noalign{\vskip 0.5mm}

    \bottomrule
    \end{tabular}
    }
    \end{table*}

\begin{table*}[ht]
    \caption{The mean EER scores of fine-tuned \textbf{with} a specific language with the data split procedure described in Section~\ref{sec:experimental}.}
    \label{tab: Fine-tuned with app}
    \centering
    \resizebox{\textwidth}{!}{
    \begin{tabular}{c|c|cccccccc}
    \hline
    \multirow{2}{*}{\textbf{Model}} & \multirow{2}{*}{\textbf{\shortstack{Fine-tuned \\ with}}} & \multicolumn{8}{c}{\textbf{Languages}} \\
    \cline{3-10}
 &  & de & en & es & fr & it & pl & ru & uk \\
    \hline
\multirow{8}{*}{RawGAT-ST}
& de & 48.61 $\pm$ 6.53 & 52.45 $\pm$ 10.74 & 49.51 $\pm$ 4.06 & 45.32 $\pm$ 11.54 & 47.33 $\pm$ 15.91 & 47.81 $\pm$ 2.12 & 49.03 $\pm$ 9.52 & 51.04 $\pm$ 5.89 \\ 
& en & 46.65 $\pm$ 6.03 & 35.35 $\pm$ 3.29 & 45.11 $\pm$ 5.11 & 42.01 $\pm$ 12.08 & 41.27 $\pm$ 14.44 & 45.68 $\pm$ 7.57 & 48.06 $\pm$ 4.37 & 43.90 $\pm$ 4.70 \\ 
& es & 54.11 $\pm$ 14.84 & 56.92 $\pm$ 13.41 & 43.79 $\pm$ 9.29 & 51.05 $\pm$ 20.87 & 49.16 $\pm$ 18.67 & 51.14 $\pm$ 19.09 & 47.85 $\pm$ 9.40 & 41.40 $\pm$ 12.45 \\ 
& fr & 45.84 $\pm$ 6.34 & 47.28 $\pm$ 0.48 & 49.56 $\pm$ 0.84 & 46.25 $\pm$ 9.77 & 47.66 $\pm$ 12.59 & 45.46 $\pm$ 8.90 & 46.01 $\pm$ 9.93 & 39.09 $\pm$ 0.70 \\ 
& it & 41.30 $\pm$ 7.22 & 44.56 $\pm$ 4.15 & 38.72 $\pm$ 2.12 & 42.35 $\pm$ 10.60 & 39.67 $\pm$ 10.17 & 44.16 $\pm$ 3.94 & 40.99 $\pm$ 14.75 & 39.75 $\pm$ 15.76 \\ 
& pl & 41.16 $\pm$ 0.60 & 26.49 $\pm$ 16.40 & 47.83 $\pm$ 7.36 & 43.82 $\pm$ 0.59 & 44.09 $\pm$ 4.48 & 40.25 $\pm$ 3.44 & 47.77 $\pm$ 13.70 & 58.93 $\pm$ 10.43 \\ 
& ru & 49.86 $\pm$ 5.95 & 46.31 $\pm$ 5.93 & 36.53 $\pm$ 3.74 & 45.02 $\pm$ 15.46 & 35.56 $\pm$ 18.29 & 51.82 $\pm$ 0.94 & 22.38 $\pm$ 3.22 & 23.19 $\pm$ 4.15 \\ 
& uk & 43.14 $\pm$ 3.21 & 53.74 $\pm$ 7.27 & 40.79 $\pm$ 3.24 & 42.36 $\pm$ 5.86 & 41.47 $\pm$ 4.57 & 43.32 $\pm$ 1.94 & 33.36 $\pm$ 1.25 & 37.92 $\pm$ 5.01 \\ 
[0.5mm]\hline\noalign{\vskip 0.5mm}
\multirow{8}{*}{Whisper+AASIST}
& de & 46.85 $\pm$ 1.42 & 39.95 $\pm$ 1.71 & 45.74 $\pm$ 3.79 & 41.81 $\pm$ 6.28 & 36.80 $\pm$ 12.11 & 46.11 $\pm$ 4.15 & 38.16 $\pm$ 8.95 & 35.83 $\pm$ 9.76 \\ 
& en & 42.51 $\pm$ 0.77 & 32.28 $\pm$ 1.97 & 43.41 $\pm$ 1.65 & 40.84 $\pm$ 4.81 & 34.24 $\pm$ 10.64 & 43.13 $\pm$ 1.39 & 35.71 $\pm$ 5.39 & 34.82 $\pm$ 10.98 \\ 
& es & 46.60 $\pm$ 2.12 & 42.59 $\pm$ 0.79 & 46.66 $\pm$ 3.87 & 43.62 $\pm$ 6.28 & 37.67 $\pm$ 12.97 & 47.49 $\pm$ 5.29 & 38.71 $\pm$ 8.82 & 36.08 $\pm$ 9.80 \\ 
& fr & 44.09 $\pm$ 0.43 & 40.89 $\pm$ 1.92 & 43.66 $\pm$ 3.10 & 39.52 $\pm$ 5.38 & 35.61 $\pm$ 11.58 & 43.87 $\pm$ 1.91 & 35.05 $\pm$ 5.94 & 33.56 $\pm$ 10.80 \\ 
& it & 44.20 $\pm$ 0.66 & 39.63 $\pm$ 0.46 & 45.13 $\pm$ 2.01 & 41.79 $\pm$ 5.31 & 36.02 $\pm$ 10.92 & 45.89 $\pm$ 1.37 & 36.90 $\pm$ 5.85 & 35.46 $\pm$ 11.50 \\ 
& pl & 46.89 $\pm$ 1.79 & 41.69 $\pm$ 4.06 & 47.89 $\pm$ 2.84 & 42.25 $\pm$ 8.08 & 38.26 $\pm$ 14.36 & 46.54 $\pm$ 4.59 & 39.32 $\pm$ 10.17 & 36.55 $\pm$ 11.59 \\ 
& ru & 44.41 $\pm$ 0.30 & 40.17 $\pm$ 1.12 & 43.90 $\pm$ 3.24 & 39.23 $\pm$ 7.79 & 36.17 $\pm$ 12.51 & 45.09 $\pm$ 3.45 & 33.89 $\pm$ 7.71 & 32.58 $\pm$ 9.97 \\ 
& uk & 44.90 $\pm$ 3.58 & 43.16 $\pm$ 0.56 & 45.20 $\pm$ 6.05 & 40.72 $\pm$ 8.34 & 37.61 $\pm$ 13.55 & 44.86 $\pm$ 8.34 & 36.81 $\pm$ 11.12 & 30.26 $\pm$ 6.00 \\ 
[0.5mm]\hline\noalign{\vskip 0.5mm}

    \bottomrule
    \end{tabular}
    }
    \end{table*}

\begin{table*}[ht]
    \caption{The mean EER scores of  fine-tuned \textbf{without} a single language.}
    \label{tab: Fine-tuned-without app} 
    \centering
    \resizebox{\textwidth}{!}{
    \begin{tabular}{c|c|cccccccc}
    \hline
    \multirow{2}{*}{\textbf{Model}} & \multirow{2}{*}{\textbf{\shortstack{Fine-tuned \\ without}}} & \multicolumn{8}{c}{\textbf{Languages}} \\
    \cline{3
    -10}
 &  & de & en & es & fr & it & pl & ru & uk \\
    \hline
    \midrule
\multirow{8}{*}{RawGAT-ST}
& de & 38.58 $\pm$ 7.93 & 46.93 $\pm$ 9.90 & 36.21 $\pm$ 1.20 & 36.94 $\pm$ 8.58 & 33.60 $\pm$ 7.11 & 29.66 $\pm$ 1.59 & 31.04 $\pm$ 1.66 & 19.90 $\pm$ 9.52 \\ 
& en & 37.62 $\pm$ 6.24 & 51.36 $\pm$ 5.58 & 37.09 $\pm$ 0.40 & 38.51 $\pm$ 7.52 & 34.50 $\pm$ 5.70 & 31.06 $\pm$ 1.14 & 35.71 $\pm$ 5.82 & 25.21 $\pm$ 15.50 \\ 
& es & 29.37 $\pm$ 0.17 & 54.01 $\pm$ 0.18 & 37.19 $\pm$ 0.29 & 28.00 $\pm$ 0.31 & 25.14 $\pm$ 0.19 & 27.66 $\pm$ 0.21 & 25.59 $\pm$ 0.20 & 22.82 $\pm$ 0.23 \\ 
& fr & 36.29 $\pm$ 7.19 & 42.81 $\pm$ 10.04 & 35.23 $\pm$ 0.35 & 37.28 $\pm$ 6.84 & 32.29 $\pm$ 6.78 & 28.74 $\pm$ 1.52 & 36.75 $\pm$ 0.60 & 22.30 $\pm$ 9.00 \\ 
& it & 40.85 $\pm$ 8.32 & 47.11 $\pm$ 5.69 & 38.42 $\pm$ 1.15 & 42.04 $\pm$ 9.62 & 36.21 $\pm$ 6.54 & 33.93 $\pm$ 1.61 & 36.40 $\pm$ 1.46 & 20.11 $\pm$ 11.56 \\ 
& pl & 40.17 $\pm$ 8.68 & 50.31 $\pm$ 7.09 & 37.56 $\pm$ 4.42 & 38.74 $\pm$ 9.91 & 35.47 $\pm$ 7.82 & 40.27 $\pm$ 7.74 & 35.92 $\pm$ 4.40 & 20.26 $\pm$ 9.85 \\ 
& ru & 40.89 $\pm$ 5.58 & 49.67 $\pm$ 8.83 & 36.51 $\pm$ 1.68 & 39.68 $\pm$ 6.49 & 36.48 $\pm$ 5.10 & 37.88 $\pm$ 1.55 & 41.49 $\pm$ 6.33 & 25.23 $\pm$ 8.83 \\ 
& uk & 35.82 $\pm$ 7.66 & 45.03 $\pm$ 10.93 & 36.10 $\pm$ 0.44 & 40.27 $\pm$ 6.52 & 31.24 $\pm$ 5.99 & 26.36 $\pm$ 0.77 & 35.46 $\pm$ 0.59 & 27.11 $\pm$ 14.97 \\ 
[0.5mm]\hline\noalign{\vskip 0.5mm}
\multirow{8}{*}{Whisper+AASIST}
& de & 42.97 $\pm$ 0.59 & 32.95 $\pm$ 1.83 & 43.70 $\pm$ 4.20 & 38.39 $\pm$ 5.84 & 35.97 $\pm$ 9.60 & 45.63 $\pm$ 0.90 & 35.29 $\pm$ 6.54 & 29.44 $\pm$ 11.26 \\ 
& en & 45.05 $\pm$ 1.03 & 36.83 $\pm$ 2.30 & 45.08 $\pm$ 5.32 & 39.02 $\pm$ 7.38 & 37.00 $\pm$ 11.23 & 46.35 $\pm$ 3.33 & 36.21 $\pm$ 8.62 & 30.06 $\pm$ 10.76 \\ 
& es & 43.13 $\pm$ 0.19 & 31.66 $\pm$ 0.97 & 44.47 $\pm$ 3.06 & 38.13 $\pm$ 6.48 & 35.57 $\pm$ 9.97 & 44.18 $\pm$ 0.95 & 35.23 $\pm$ 7.85 & 29.30 $\pm$ 10.50 \\ 
& fr & 44.25 $\pm$ 0.45 & 32.29 $\pm$ 0.99 & 44.85 $\pm$ 4.52 & 40.31 $\pm$ 7.14 & 36.30 $\pm$ 10.62 & 45.12 $\pm$ 1.91 & 36.21 $\pm$ 8.19 & 29.80 $\pm$ 10.30 \\ 
& it & 43.55 $\pm$ 0.27 & 33.11 $\pm$ 0.71 & 43.96 $\pm$ 4.92 & 38.33 $\pm$ 6.95 & 37.09 $\pm$ 11.02 & 44.74 $\pm$ 1.94 & 35.36 $\pm$ 7.70 & 29.65 $\pm$ 10.61 \\ 
& pl & 43.14 $\pm$ 0.34 & 32.21 $\pm$ 2.39 & 43.69 $\pm$ 5.03 & 37.18 $\pm$ 6.02 & 35.85 $\pm$ 10.03 & 45.30 $\pm$ 1.27 & 34.30 $\pm$ 7.11 & 29.56 $\pm$ 11.15 \\ 
& ru & 43.23 $\pm$ 0.63 & 32.04 $\pm$ 1.22 & 44.62 $\pm$ 4.47 & 38.74 $\pm$ 6.22 & 36.49 $\pm$ 10.27 & 45.06 $\pm$ 1.30 & 36.89 $\pm$ 7.93 & 29.84 $\pm$ 11.01 \\ 
& uk & 43.55 $\pm$ 0.28 & 32.51 $\pm$ 1.64 & 44.80 $\pm$ 3.60 & 38.44 $\pm$ 5.35 & 36.34 $\pm$ 9.40 & 45.89 $\pm$ 0.61 & 36.06 $\pm$ 6.16 & 31.77 $\pm$ 12.43 \\ 
[0.5mm]\hline\noalign{\vskip 0.5mm}
    \bottomrule
    \end{tabular}
    }
    \end{table*}

\end{document}